\documentclass{basi}
\usepackage{epsfig,graphicx} 
\begin{document}
\title[Quest for solar gravity modes]{The quest for solar gravity modes: probing the solar interior} 
\author[S. Mathur]%
       {Savita Mathur \thanks{e-mail:smathur@iiap.res.in} \\ 
        Indian Institute of Astrophysics, Bangalore 560034}
\maketitle
\label{firstpage}
\begin{abstract}
 The solar gravity modes are the best probes to improve our knowledge on the solar interior, as they spend most of their time in the radiative zone, which represents 98\% of the solar mass. Many attempts have been led to observe them using different techniques: either individually, then adding some statistical approach or more recently, globally  leading to the detection of the signature of asymptotical properties of these modes. Then, several theoretical works have been done to quantify the effect of detecting g-mode on solar modeling and on the rotation profile. We will give here an update on the g-mode detection. Then, we will study an example of a theoretical work showing how their detection would improve our knowledge on the dynamics of the solar core as well as an application on the detection of the global properties to infer some physical inputs in solar models. 

\end{abstract}


\section{Introduction}
\label{sec:intro}
During the last decades, helioseismology has proved to be the best tool to improve our knowledge on the Sun and to give constraints on solar modelling. Among all the discoveries made thanks to helioseismology, it enabled to have, with a good precision, the position of convection zone (Corbard et al. 1998), the content of He (Basu \& Christensen-Dalsgaard 1997), and to correct the discrepancy between the prediction and the measurements of neutrinos fuxes (Turck-Chi\`eze et al. 2001). It also contributed to obtain the sound speed profile. Finally, the solar rotation profile was obtained by inverting helioseismic data down to $\sim$~0.3~R$_\odot$ (Thompson et al 2003, Couvidat et al. 2003a, Garc\'ia et al. 2004). However, we still need to find and put more constraints on the solar models as the rotation rate is not known in the core and discrepancies between the determination of the sound speed profiles from these models with the new surface abundances of Asplund et al. (2005) and helioseismic observations (Turck-Chi\`eze et al. 2004a, Antia \& Basu 2005) are still present. We need to improve our knowledge on the dynamics and the structure of the solar core. Today, only the acoustic (p) modes have been detected and studied. But to have more information on the solar core, we need to detect the gravity (g) modes that are trapped in the radiative zone and thus are very sensitive to this region of the Sun. Being evanescent in the convective zone, they are expected to have low amplitudes at the surface of the Sun (Belkacem et al. 2009). This quest started a few decades ago but did not lead to a non-ambiguous detection of g modes (Garc\'ia et al. 2001, Gabriel et al. 2002, Turck-Chi\`eze et al. 2004b). The new generation of instruments such as PICARD or GOLF-NG (Turck-Chi\`eze et al. 2008) are dedicated to this search. First, we explain the different techniques used to search for the g modes (individual and global detections) and we give the latest status of their detection. Then we describe a theoretical application of the detection of individual g modes on the rotation profile. Finally, we show how the detection of their global properties can help us in constraining some physical inputs of the solar models.

\section{Looking for gravity modes}


\subsection{Individual search}

\begin{figure} [h!] 
  \centering
   \includegraphics[height=4.5cm]{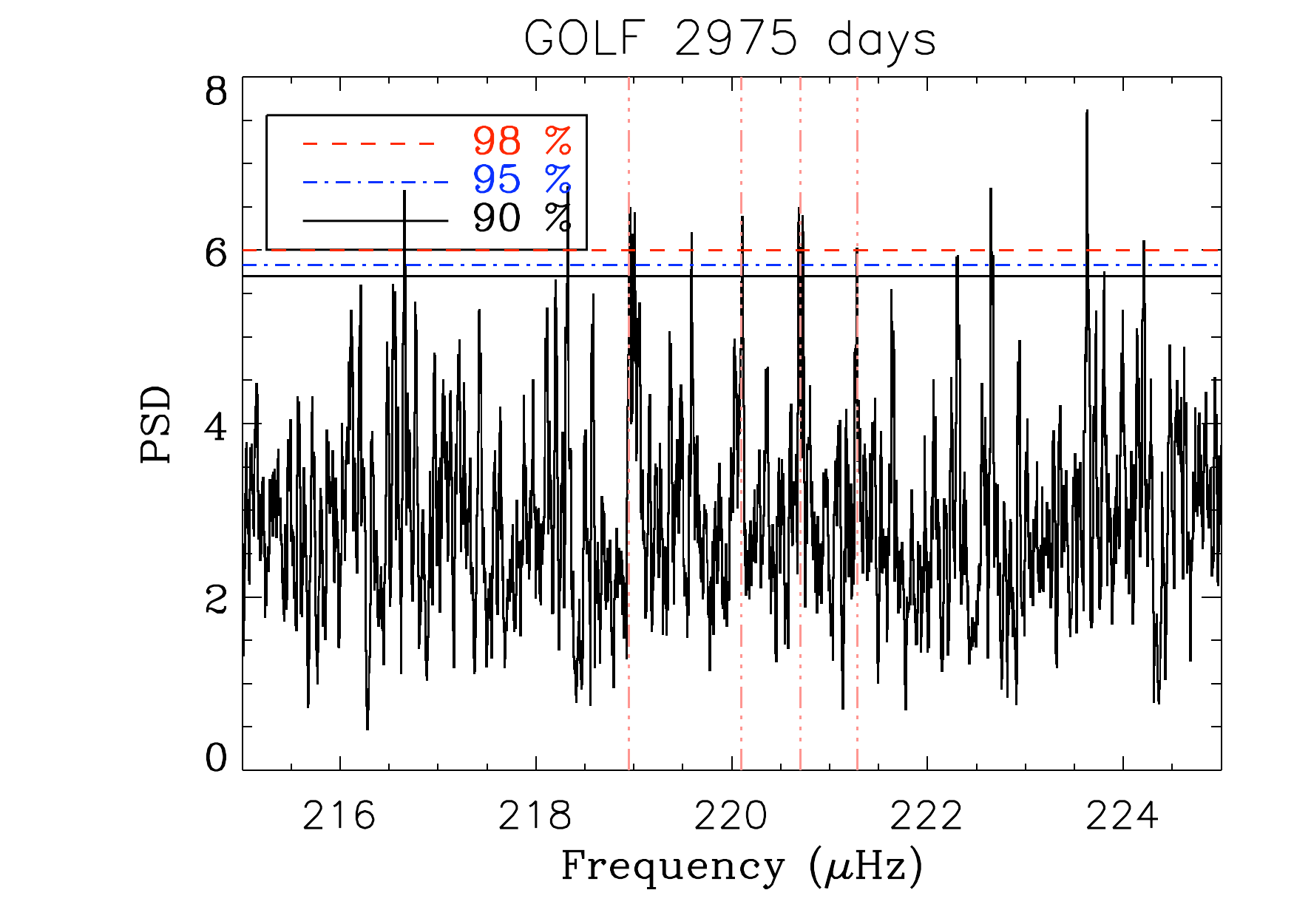}

  \caption[]{The g-mode candidate around 220~$\mu$Hz using 2975 days of GOLF data with the different confidence levels of 90, 95 and 98 \%.(from Mathur et al. 2007). \label{mathur-fig:001}

}\end{figure}

The first method to look for g modes consists in looking for individual modes as it is done for the p modes. However, because of a very low signal-to-noise ratio (due mainly to the high solar granulation noise in this frequency range), statistical approaches have been developed (Broomhall et al. 2007 and references therein) to calculate the threshold for a given confidence level by using Monte Carlo simulations. It led to the detection of a g-mode candidate, using almost 3000 days of GOLF\footnote{Global Oscillations at Low Frequency (Gabriel et al. 1995)} data. This candidate has more than 98\% of confidence level not to be due to noise (Fig.~\ref{mathur-fig:001}). Besides, one theoretical work done by Cox \& Guzik (2004) showed that the g mode around 220~$\mu$Hz has the hightest predicted amplitude. As we cannot identify this candidate in terms of its $\ell$, $n$ and $m$, several scenarios have been proposed to explain it, thus giving different values of splittings (Turck-Chi\`eze et al. 2004b, Mathur et al. 2007).  Moreover, this candidate seems to evolve with time in the data. Finally, one of the component of this candidate at 220.7~$\mu$Hz has also been observed in VIRGO\footnote{Variability of solar IRradiance and Gravity Oscillation (Fr\"ohlich et al. 1995)} at the same frequency and that was followed with time and appeared to be stable (Jimenez \& Garc\'ia 2009).

\subsection{Looking for global properties}

The second method to look for g modes is based on their asymptotic properties. For $n \gg \ell$, g modes are periodically spaced in periods. We study $\Delta P_\ell$, the difference between two consecutive orders, for a given $\ell$: $P_{\ell, n+1} - P_{\ell, n}$. Using different solar models, it has been shown that for $\ell$ = 1, it is well constrained between 24 and 25 minutes (Mathur et al. 2007). By calculating the PS2, which is Fourier Transform of the Power Spectrum (PS1) with a Sine-wave fitting, we looked for a pattern around this value in the PS2. 

\begin{figure} [htb] 
  \centering
  \includegraphics[height=5.5cm]{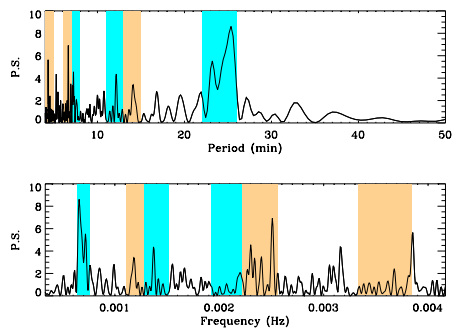}
  
  \caption[]{PS2 as a function of the period (top) and as a function of the frequency (bottom). Blue areas correspond to the expected regions of $\Delta P_1$ and its first harmonics whereas orange areas correspond to expected regions of $\Delta P_2$ and its first harmonics (from Garc\'ia et al. 2008a). \label{mathur-fig:002}

}\end{figure}

It led to the detection of a tall and wide structure around the theoretical period with 9.5 years of GOLF data and having a confidence level higher than 99.5\% (Garc\'ia et al. 2007). Then, adding 25\% of data obtained during the minimum of solar activity, this structure is even taller and is the most prominent peak in the PS2 (Fig.~\ref{mathur-fig:002}).\\
Then, we used the phase information of the sine-wave fitting to retrieve some information on the rotation rate in the solar core. By filtering the PS2 around $\Delta P_1$ and its first harmonic, we reconstructed the signal of the PS1 (see Garc\'ia et al. 2007 for more details). This technique is very powerful as with the PS2 we could not distinguish between two rotation rates in the core. Now we are able to recover the positions of the peaks in the PS1. We used different rotation profiles to reconstruct the signal with one solar model. Then, we calculated the correlation rates between these reconstructed signals and the one from GOLF data. The highest correlation rate tends to favor a core rotating faster than the rest of the radiative zone.

\section{Some applications of the detection of gravity modes}

\subsection{Inferring the rotation profile with individual detection}

\begin{figure} [!h] 
  \centering
  \begin{tabular}{p{6.3cm}p{6.3cm}}
  	\includegraphics[width=6.3cm]{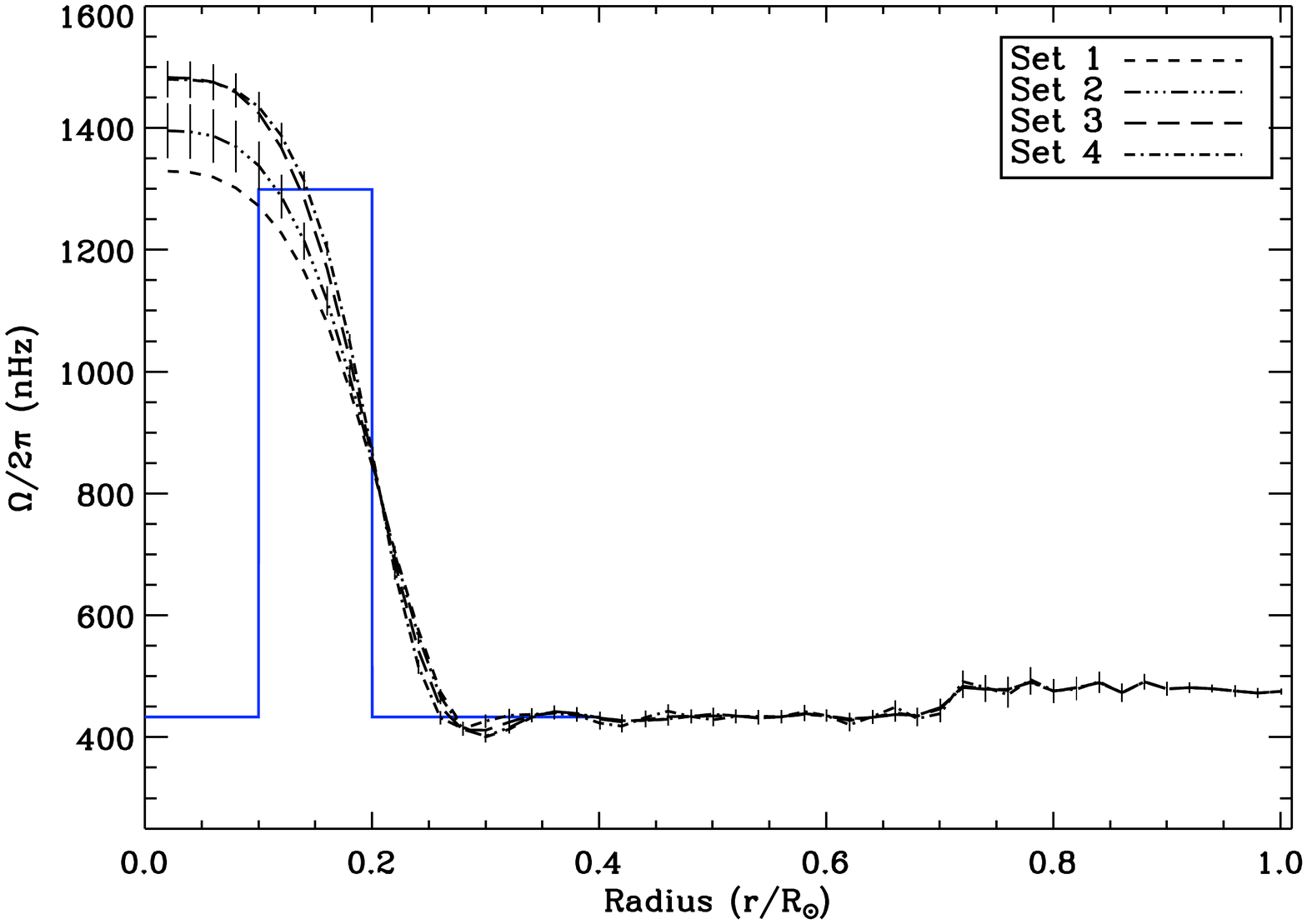} &
	\includegraphics[width=6.3cm]{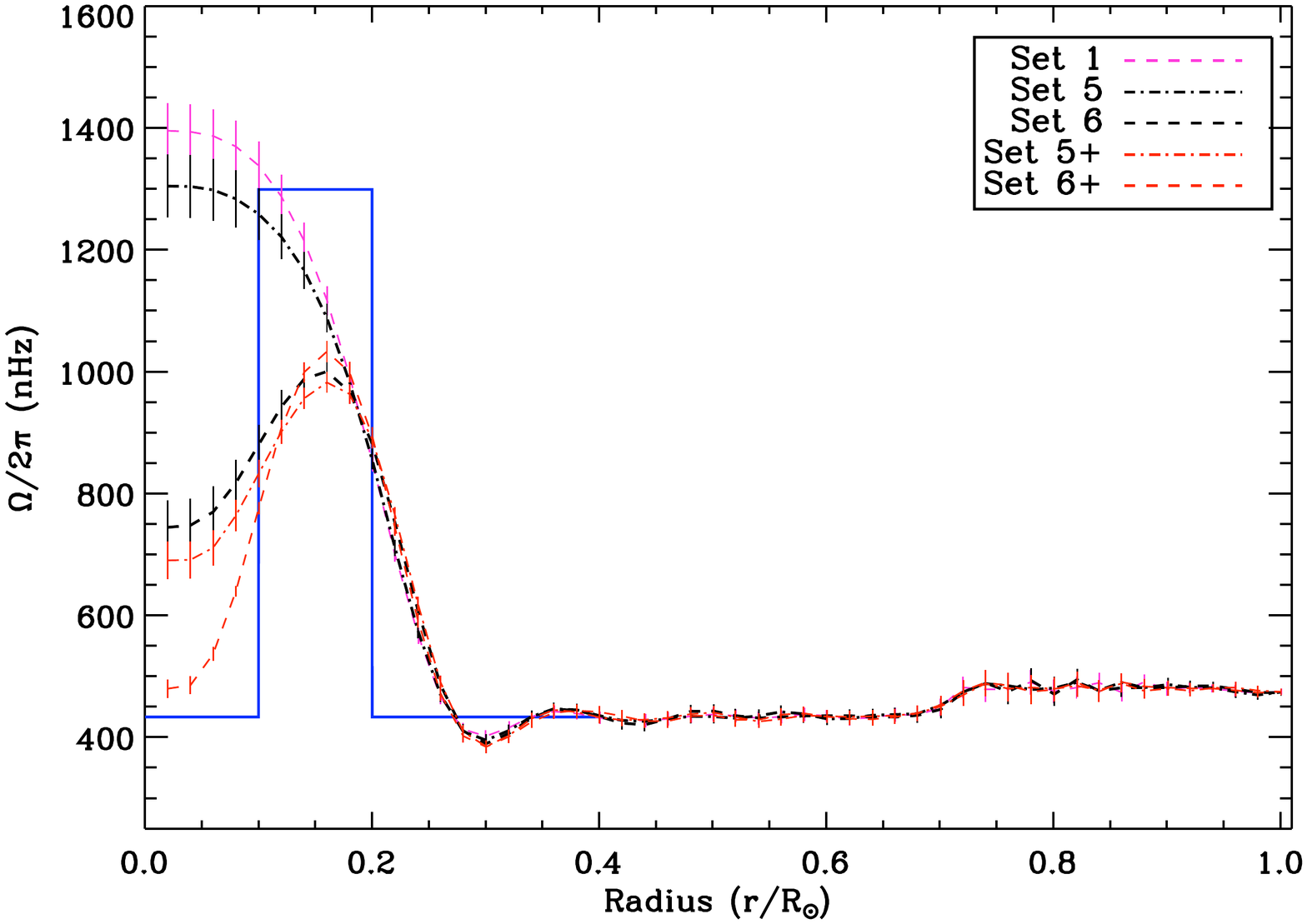}
	\end{tabular}
  
  \caption[]{Equatorial rotation profiles inferred from inversions. Left panel: inversions with only p modes (sets 1, 2, 3 and 4). Right panel: adding g modes (sets 5, 6, 5+ and 6+).\label{mathur-fig:003}

}\end{figure}

In this section, we wanted to see how low-degree high-frequency p modes as well as g modes can improve our knowledge on the deep solar interior. We applied an inversion code based on the Regularized Least-Squared methodology (Eff-Darwich \& Hern\'andez 1997). We simulated one artificial rotation profile (blue curve in Fig.~\ref{mathur-fig:003}) to calculate artificial data of p and g modes. We used 8 different sets of modes with present error bars (Garc\'ia et al. 2008). \\

Sets 1 to 4 contain only p modes. The modes $\ell \ge$4 go up to 3.9~mHz. Whereas for the modes $\ell \le$3, in set 1, they are taken only below 2.3~mHz, in set 2, below 3.9~mHz. Then in sets 3 and 4, the uncertainties of these modes below 3.9~mHz are respectively divided by 3 and 6. Figure~\ref{mathur-fig:003} shows that by decreasing the error bars of these low-degree high frequency p modes, the results are closer to the rotation rate between 0.1 and 0.2~R$_\odot$. These modes are indeed sensitive to this region if we look at their inner turning points (see Garc\'ia et al. 2008b, Mathur et al. 2009).\\
Then, in sets 5 and 6, we add respectively one (the candidate $\ell$ = 2, $n$ = -3) and 20 g modes ($\ell$ = 1 and 2, $n$ = -1 to -10). These low-degree modes are the most likely to be observed in the future. They are associated to two error bars: 75~nHz and 10~nHz (index +). We can see that with one g mode and a large error bar, we do not retrieve the rate in the core. But by increasing the number of g modes and decreasing their error bars, we start to see the decrease below 0.1~R$_\odot$. We almost reach the rate in the core. This means that we need a few tens of g modes and reasonable error bars to have some information on the rotation rate in the solar core, while for the rotation profile between 0.1 and 0.2~R$_\odot$, we need to decrease the uncertainties of the low-degree high-frequency p modes.
Finally, using real data, we are confident on the profile down to $\sim$0.17R$_\odot$ using several splittings for the g-mode candidate (Mathur et al. 2008).

\subsection{On the physical processes with global properties}

In this section, we use the detection of the asymptotic properties of g modes. As explained in Section~2.2, we reconstruct the signal of the PS1 for different solar models and several artificial rotation profiles. The solar models differ from several physical inputs like the microscopic diffusion, diffusion in the tachocline, surface abundances. Model S is the standard model of Christensen-Dalsgaard et al. (1996), the seismic model is described in Couvidat et al. (2003b) and for the other models, their description is given in Garc\'ia et al. (2008c). Concerning the rotation profiles, they are defined by the rotation rate in the core ($\Omega_c$) and the limit of the core in solar radius ($r_i$ = 0.1, 0.15 and 0.2 R$_\odot$).\\
First, we notice that the correlation rates are higher for the models using the abundances of Grevesse \& Noels (1993) compared to the rates for the models without microscopic diffusion or with the abundances of Asplund et al. (2005). This is compatible with the results using only p modes to calculate the sound speed profile as we observe higher discrepancies between these latter models and the observations. Let us note that the highest correlation rates are obtained when the core is rotating faster in average than the rest of the radiative zone.
\begin{figure}
  \centering
 
  \includegraphics[height=4.3cm]{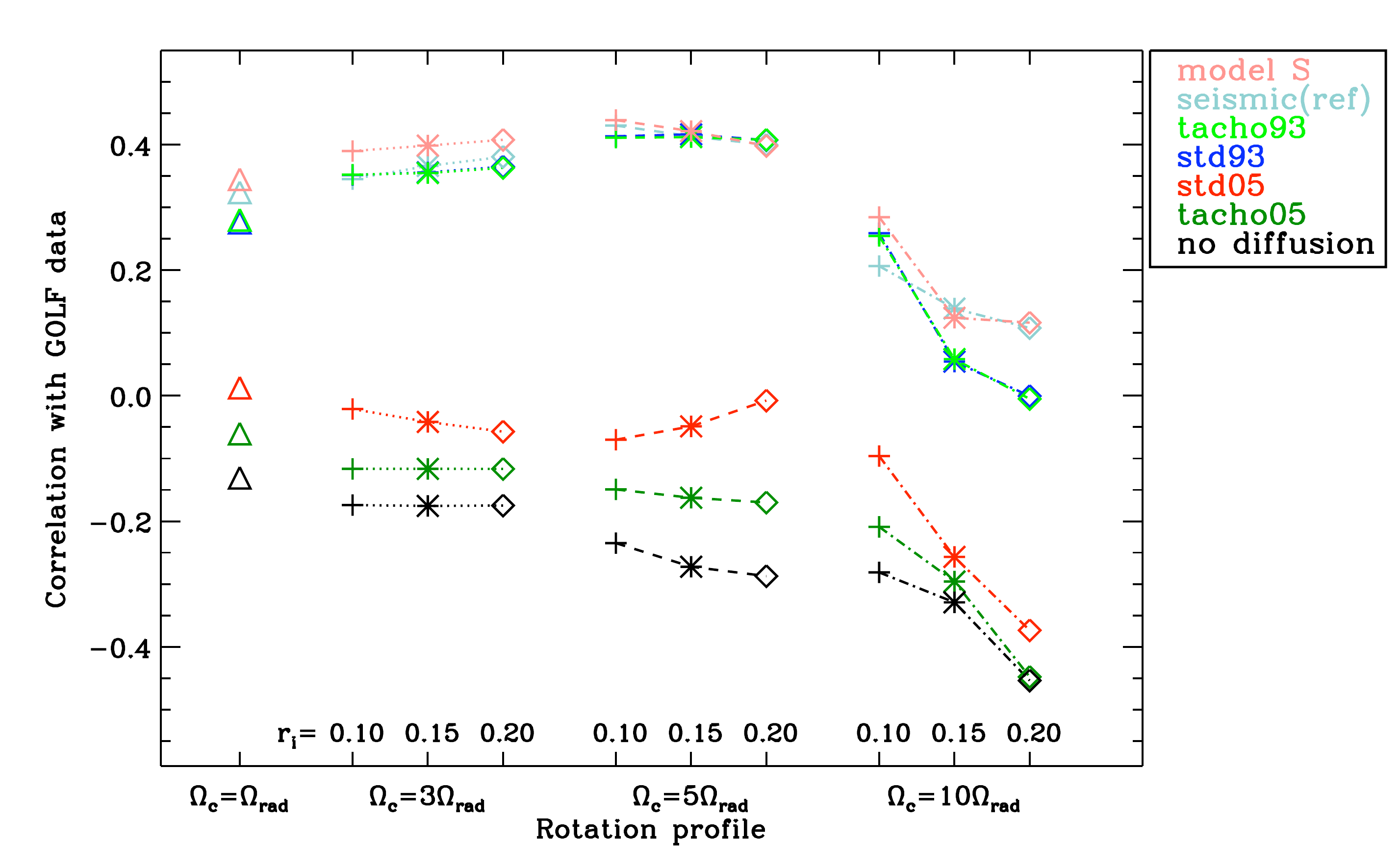}

  \caption[]{Correlation rates between the reconstructed signals from GOLF and for different solar models as a function of the rotation profiles (from Garc\'ia et al. 2008). \label{mathur-fig:004}

}\end{figure}

\section{Conclusions}

A signal fully compatible with the $\ell$ = 1 g modes has been detected and confirmed with more than 11 years of GOLF data. This detection is also compatible with the p-mode observations. One g-mode candidate was detected with more than 98~\% of confidence level. One of its component has been studied with other instruments and seems to be stable with time. We saw that to reconstruct the rotation profile below 0.2~R$_\odot$, we need both low-degree high-frequency p modes and a few tens of g modes with a good accuracy. \\


\section*{Acknowledgements}

This work has been partially supported by the CNES/GOLF grant.

\label{lastpage}
\end{document}